\documentclass[12pt]{spieman}  
\usepackage{amsmath,amsfonts,amssymb}
\usepackage{graphicx}
\usepackage{setspace}
\usepackage{tocloft}
\usepackage[symbol]{footmisc}

\title{Toward precision radial velocity measurements using Echelle spectrograph at Vainu Bappu Telescope}

\author[a]{Sireesha Chamarthi}
\author[a]{Ravinder K. Banyal}
\author[a]{S. Sriram}
\affil[a]{Indian Institute of Astrophysics, Bangalore, India}

\cftpagenumbersoff{figure}
\cftpagenumbersoff{table} 
\begin{document} 
\maketitle

\begin{abstract}
The Echelle spectrograph operating at Vainu Bappu Telescope (VBT) is a general purpose instrument designed for high resolution spectroscopy. It is being considered for precision Doppler measurements without altering the existing design and basic usage. However, the design level limitations and environmental perturbations are a major source of instability and systematic errors. As a result, a small Doppler signal in the stellar spectra is completely swamped by the large and uncontrolled instrumental drift. In this paper, we discuss some of the remedial measures we took to improve the radial velocity performance of the spectrograph. We show that a new auto-guider assembly has greatly reduced the mechanical jitter of the star image at the fibre input, making the illumination of the spectrograph slit at the other end stable. We have also installed an iodine absorption cell to track and eliminate the instrumental drifts to facilitate precision radial velocity observations. Furthermore, we have developed a generic algorithm that uses iodine exposures to extract the stellar radial velocities without the need for the complex forward modeling. Our algorithm is not accurate to the level of traditional iodine technique. However it is convenient to use on a low-cost general-purpose spectrograph targeting a moderate Radial Velocity (RV) precision at a few 10-100~$\textrm{ms}^{-1}$ level. Finally, we have demonstrated the usefulness of our approach by measuring the RV signal of a well known short-period, planet-hosting star.     
\end{abstract}


{\noindent \footnotesize\textbf{*}Sireesha Chamarthi,  \linkable{sireesha@iiap.res.in} }

\begin{spacing}{2}   

\section{Introduction}
\label{sect:intro}  

Discovery of a planet around a main sequence star, 51 Peg, in 1995 marked the beginning of exoplanet science \cite{first_exo}. With the advancement in technology and sophisticated detection techniques, the field has rapidly expanded to new research areas. Exoplanets are detected and characterize by many methods such as transit observations, radial velocity method, microlensing, direct imaging, and astrometry etc. However, none of these techniques can independently  determine all the properties of the exoplanets and host stars \cite{detection}. This inadequacy is often addressed by taking a complementary approach. For instance, planetary size is obtained from transit observations  and mass ($m\cdot \sin i$) is obtained from the RV measurements. These two parameters are then used for inferring the bulk density of the planet, which helps further classifying the the planet as rocky or gaseous.

Increase in the ground as well as space-based transit surveys such as the XO project \cite{XO}, Transatlantic Exoplanet Survey \cite{Tres}, Kepler mission \cite{kepler}, COROT space mission \cite{COROT}, Transiting Exoplanet Survey Satellite (TESS) \cite{TESS} etc have enormously contributed  to the overall count and the rate at which new exoplanet discoveries are made. Transit method is a more accessible technique for exoplanet detection. It detects the periodic dips in the light curve of the stars for confirmation of the orbital companions. The transit measurements give mainly the following parameters: transit depth, duration, the duration of ingress/egress, and the orbital period of the exoplanet. Even though the method involves an uncomplicated detection technique, there is a chance of very high false positives \cite{false_positive}. Thus, transit technique requires follow-up RV observations to confirm the orbital companion. Also, RV characterization of the planets is to be followed up with ground-based telescopes to obtain parameters such as orbital period, eccentricity, minimum mass of the planet. Therefore, in light of rapid exoplanet discoveries ushered by the current and the future transit missions, the RV follow-up studies from the ground has assumed paramount importance.

A major contribution to precision RV measurements using ground-based telescopes came from stabilized spectrographs using simultaneous Th-Ar arc lamp for wavelength calibration, e.g. HARPS \cite{Pepe} and the iodine absorption cell technique developed by Marcy and Butler \cite{marcy}. The current state-of-the-art instrumentation has enabled the RV measurements down to 1~$\textrm{ms}^{-1}$ level \cite{onems}. This is achieved with ultra-stable spectrographs combined with advanced data reduction and analysis algorithms.  Building and operating a dedicated high precision spectrograph for RV studies is still uncommon. Further, the telescope time at these high-end facilities is oversubscribed. A dedicated RV spectrograph  may not even be a preferred option for many observatories focusing on various other science cases and legacy programmes. However, already established and operational small telescopes with high-resolution spectrographs can be easily adapted to address some exoplanet science cases such as a search for short period giant planets, RV follow-up and characterization of hot-Jupiters detected around bright stars, the study of pulsating stars \cite{twom_class,RV_followup}.  In this scenario, it is appropriate to utilize the existing spectrographs operating on small telescopes with moderate RV precision of 10 -100 $\textrm{ms}^{-1} $~ for science cases such as \cite{onem_class}: (i) confirmation of the presence of hot Jupiter exoplanets (ii) refining orbital parameters of the companions  (iii) verification of false positives in transit measurements (iv) study of binary or pulsating stars. Importance of using a network of small telescopes for Doppler studies have been discussed by D.~Mkrtichian {\it et al} (2008) \cite{twom_class}. Normally, to deploy the existing spectrographs for high precision RV studies, either the spectrograph is to be vacuum stabilized or iodine absorption cell has to be used with complex data processing pipeline. This requires technical expertise that may not be available all the time.  

 In this paper, we describe our efforts to utilize the echelle spectrograph operating at the Vainu Bappu Telescope, Kavalur for precision RV measurements. To enhance the science capabilities of the spectrograph, an iodine absorption cell is incorporated at the entrance slit of the spectrograph. We propose a simple algorithm as an alternative to the forward modeling approach, to detect the instrumental shifts during the observations by utilizing the dense forest of absorption lines obtained from a temperature stabilized iodine cell. Apart from this, an auto-guider is installed at the prime focus of the telescope to direct the star-light to the entrance slit of the spectrograph via optical fibre. The spectrograph has design level limitations like movable grating and tilt in spectra. With the movable components and the inherent spectral tilt, the standard forward modeling of Iodine and star observations used for extracting the Doppler parameter becomes more complicated. Our proposed approach does not use forward modeling but aims to track and eliminate the instrumental shifts in the spectrograph using iodine cell absorption lines. This method is similar to the technique which uses time-tagged Th-Ar frames (closest in time when the science target is observed) to minimize the drift errors \cite{super_calibration,uves}. However, unlike time-tagged Th-Ar case, the iodine absorption lines interspersed with star spectra, can track the instrumental shifts even during the observations of science target.  

A short description of the spectrograph and the upgrades related to precision RV studies are described in Section \ref{sect:spectrograph}. The observational strategy for instrumental drift tracking algorithm is presented in Section \ref{sect:procedure}. We have validated the algorithm with a well-studied planet-hosting star, tau Bootis, the results are presented in Section \ref{sect:results}. Discussion about the methodology is presented in Section \ref{sect:disc}.
In Section \ref{sect:concl} we have discussed the limitations of our approach and possible areas of improvements along with concluding remarks.

\section{Spectrograph details - Measures to improve the capabilities of the instrument }
\label{sect:spectrograph}
\subsection{Spectrograph description and limitations }
The Echelle spectrograph operating at VBT is designed to carry out high-resolution spectroscopic observations such as stellar abundance analysis, the study of pulsating variables, chemical compositions of stars, and probing stellar and galactic kinematics among many others \cite{VBTechelle}. The starlight from the telescope is inserted into a 100~$\mu$ $\textrm{m} $ multi-mode optical fibre of 45 $\textrm{m} $ length placed at the prime focus of the telescope. The light from the exit end of the fibre is fed to the input optics which focuses it onto the slit. The output of the slit is collimated and forwarded onto a prism. The prism output is given to an echelle grating for primary high dispersion. The echelle grating has a groove density of 52.67 $\textrm{grooves/mm} $. The reflective grating acting in quasi littrow mode feeds the light back into the cross-disperser prism. In the return path, the collimator acts as a camera and focuses the light onto a 4K$\times$4K CCD. It is a compact low-cost instrument, with the cross-disperser prism and the collimator in the double pass. 

The echelle grating operating at a blaze angle of (${\theta}${\tiny B}= $70\,^{\circ}$), enabling high resolution of $R=60,000$ for stellar observations. The spectrograph has a wavelength coverage from 4000-10,000 {\AA} and a limiting magnitude of  V $\leq$ 10. As it is a general purpose instrument, the long-term stability of the spectrograph is not established. In an earlier experiment, we had estimated the spectral drift about $ \pm$1 pixel ($ \pm $1000 \textrm{ms}$ ^{-1} $) drift over several nights of observations \cite{stability_VBT}. Since the instrument is not originally designed for extreme stability, the following factors could limit the achievable RV precision : 

\subsubsection{Mechanical effects}
\begin{figure}
   \begin{center}
   \begin{tabular}{c} 
   \includegraphics[scale = 0.55]{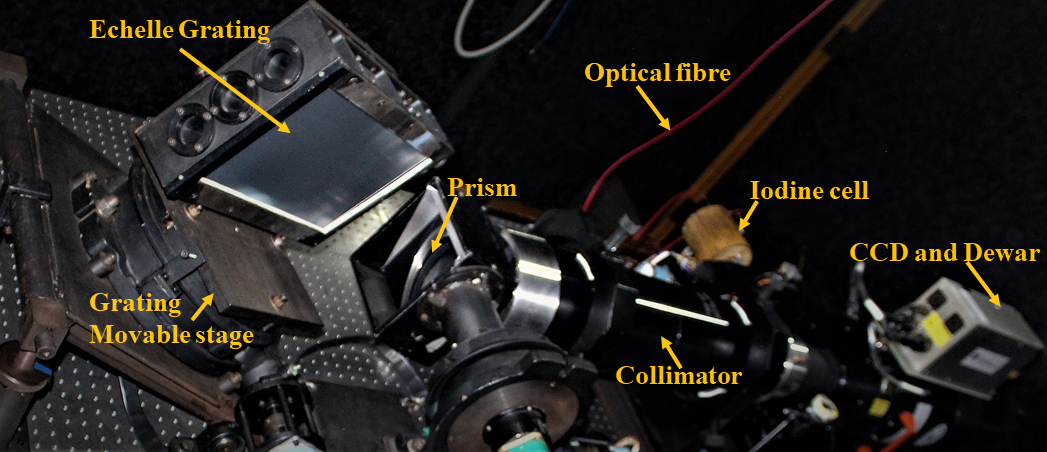}
	\end{tabular}
	\end{center}
   \caption
   { \label{fig:Echelle} 
The rear view of the spectrograph from the Echelle grating side. The grating is the primary dispersive element mounted on a movable stage to cover the wavelength gaps at the start and end of each echelle order. The star-light is fed through a 100~$\mu$ $\textrm{m} $ multimode optical fibre. The input optics converts the telescope (F/3) beam to (F/5) and focuses it onto the slit. The light from the slit is fed to an (F/5) collimator using a flat mirror oriented at 45$^\circ$ with respect to the beam. The beam from the collimator falls on to the cross-disperser prism and then to Echelle grating for dispersion. The dispersed light from the grating is forwarded to the prism again in the return path. In the second pass, the collimator acts as a camera, and the beam is focused on the 4K$\times$4K detector.}
   \end{figure}

The spectrograph has a movable grating for accommodating the wavelength gaps at the edges of Free Spectral Range (FSR) of each echelle order. Figure \ref{fig:Echelle} shows the spectrograph with the optical components along with the grating stage. Due to the movement of the optomechanical components, there is no fixed zero-point of the spectrograph. As a result of this, the Point Spread Function (PSF) modeling might carry asymmetries along with systematic errors \cite{whitepaper}. In our previous studies, we have shown that the PSF of the spectrograph varies spatially and has notable asymmetries \cite{PSF_vbt}. Further, the changes in the grating position and variations in the pupil illumination also contribute to RV errors \cite{whitepaper}.

\subsubsection{Tilt in the spectra}
The spectrograph is operated in a quasi-littrow mode of operation with $\theta= 1.1^\circ$ \cite{VBTechelle}. The prism placed in the path of the grating in the spectrograph assembly, partially disperses the light before forwarding on to the grating. The partial cross-dispersion before the grating results in the out of plane angle of the grating as a function of wavelength. This results in a tilt in the slit image at the detector as a function of wavelength \cite{tilt}. The RV precision is compromised when the tilt in the spectra is not properly taken care (modeled) by the extraction software \cite{tilt_analysis}.

Aforementioned effects are due to design constraints of the spectrograph and cannot be fully corrected unless a major instrument upgrade is considered. However, as it is a general purpose spectrograph, the instrument is used for many high-resolution spectroscopic observations. Therefore, we have only attempted addressing issues that don't affect the existing functionality of the spectrograph. Currently, the factors which fall into this category are the slit-illumination errors and non-simultaneous wavelength calibration. Addressing these issues will improve the performance of the spectrograph and enables exoplanet science with the instrument. 

  \begin{figure}
   \begin{center}
   \begin{tabular}{c} 
   \includegraphics[scale = 0.55]{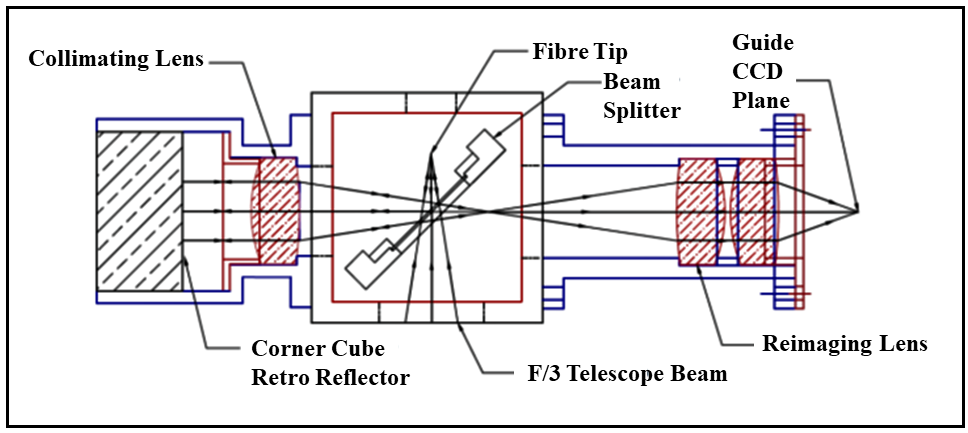}
	\end{tabular}
	\end{center}
   \caption
   { \label{fig:auto_guider} 
The schematic of the auto-guider installed at the prime cage of the telescope.  Star-light from the primary mirror of the telescope (F/3) falls on the beam splitter, that reflects 8$\%$ of the beam onto the guide CCD (through re-imaging lens) to monitor target guiding of the telescope. The fibre is re-imaged onto the guide CCD using the collimating lens and the Corner cube retro-reflector and re-imaging lens. The fibre position is estimated on the guide CCD and the guiding algorithm redirects the star-light onto the fibre co-ordinates.}
   \end{figure}   
   
\subsection{Auto-guider --correction of illumination errors}
The light beam from 2.3 m primary aperture of the telescope converges to the prime focus. A 100~$\mu\textrm{m} $ (2.7 arc-sec) optical fibre picks the star-light from the prime focus and carries it to the entrance slit of the spectrograph. The star image on the fibre entrance does not remain stable due to imperfect guiding of the telescope. The wandering motion of the source at the input-side of the fibre is also replicated at the exit-side, preventing the stable illumination of the spectrograph pupil. In the absence of scrambling, the illumination errors in multimode fibres can be a significant source of RV inaccuracies \cite{autoguider}. To reduce the impact of illumination errors, an auto-guider was developed and installed at the prime cage of the telescope. Figure \ref{fig:auto_guider} shows the schematic of the guider.

The auto-guide assembly consists of a beam splitter with 92$\%$ transmission and 8$\%$ reflection, residing close to the prime focus of the telescope. The beam splitter is placed $\sim30$~ $\textrm{mm} $ in front of the optical fibre at 45$^\circ$ angle such that the 8$\%$ reflected beam is directed towards the guide CCD and transmitted beam is directed onto the spectrograph fibre.

\subsubsection{Integration and calibration of the auto-guider}
\label{subsubsect:integration-guider}

During the assembly and calibration of the guider, the optical fibre that carries the star light to the spectrograph is illuminated from the rear-end (spectrograph side) with a light source. The optics of the guider is aligned in such a way that the fibre exit (front end) is re-imaged onto the guide camera using the corner-cube-retro-reflector as shown in Figure \ref{fig:auto_guider}. The location of the pick-up fibre on the guide CCD camera is set as a reference and frozen for the guiding purpose during the integration. During the observations, the guiding algorithm operates in closed-loop to keep the reflected star image on the reference location defined on the guide CCD during the calibration. 

The telescope has an equatorial mount, the tracking is done in the Right Ascension (RA) direction with the Declination (DEC) axis locked. The guiding algorithm controls the telescope motors and adjusts the RA and DEC such that the star centroid is well within the fibre center. The guiding algorithm starts when the star crosses $\pm$2~pixels on Guider CCD (0.54 arc-sec). The guider adjusts the telescope RA and DEC such that the star image is pulled within $\pm$1~pixels on the Guider CCD. The $\pm$2~pixels on Guider CCD acts as a blind zone to the guider, only when the star image stray outside this window, the guiding algorithm turns on to bring the star within $\pm$ 1 pixel. The green window in Figure \ref{fig:all_guide} shows the $\pm$2~pixel (0.54~arc-sec) window on the guide CCD in comparison with the optical fibre (2.7 arc-sec) displayed in magenta color.

\subsubsection{Results of auto-guider at the telescope}   
\begin{figure}
   \begin{center}
   \begin{tabular}{c} 
   \includegraphics[scale = 0.5]{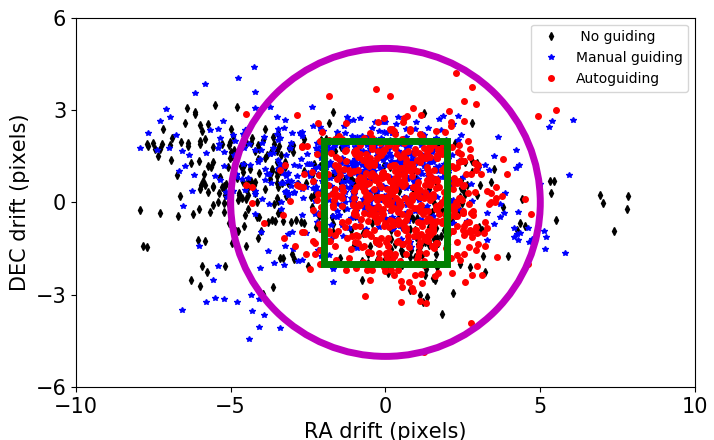}
	\end{tabular}
	\end{center}
   \caption
   { \label{fig:all_guide} 
  Comparison of star centroid scatter for \textit{no guiding}, \textit{manual guiding}, and \textit{auto-guiding} case. As the telescope has an equatorial mount the drift is seen in the RA direction. The green box corresponds to the 2 x 2 window where the auto-guider algorithm acts and brings the star centroid to within the range. The magenta circle represents the 100 $\mu $ diameter of the fibre core. In case of auto-guiding 65.2 $\%$ of the time the star falls within the square window. The star centroid falls 50.6 $\%$ and 21 $\%$ for manual and no-guiding respectively.
    }
   \end{figure}

In order to characterize the performance of the auto-guider a bright star located close to Zenith is chosen. A series of observations with \textit{auto-guider}, \textit{manual-guiding} and \textit{no-guiding} are taken with the star for a duration of 2 hours. We have compared \textit{auto-guiding} with \textit{manual-guiding} and \textit{no-guiding}. In case of \textit{manual guiding}, the reflected star centroid on the guide CCD was manually positioned on to the fibre center location.  During the \textit{no guiding case}, the telescope was tracking the star without any guiding intervention. Figure. \ref{fig:all_guide} shows the comparison of the centroid scatter of a star on the guide CCD for settings of \textit{auto-guiding, manual-guiding}, and \textit{no-guiding}. In the case of \textit{auto-guiding}, the RMS scatter of the star centroid obtained at the guide CCD is 1.61 pixel about the fibre centre. As shown in green box of Fig. 3, the$\pm$ 2 pixels window is a blind region for the guider. Due to motor movement and the telescope system tracking limitations we could not achieve a window smaller than this for guiding. The RMS scatter in the centroid of the  star image was 3.35 pixel and 2.46 pixel for \textit{no-guiding} and \textit{manual-guiding}, respectively. The improvement in the RV precision due to installation of the auto-guider is around $\pm$ 200$\textrm{ms}^{-1} $.

\subsection{Iodine absorption cell}
A Th-Ar emission lamp is used as a wavelength calibration unit at the spectrograph. The light from the calibration source is gathered by a separate 100~$\mu$m optical fibre which couples it to the spectrograph fibre in the prime cage. The calibration frames are taken before and after the stellar observations. The existing design lacks simultaneous calibration and thus we have used iodine absorption cell for tracking the instrumental shifts. 
 
Iodine absorption cell is installed at the entrance slit of the spectrograph in the telescope beam path. The iodine cell is stabilized to a temperature of 60$\pm$0.2$^\circ$C. The cell is placed on an automated translation stage that moves it in and out of the beam path during observations.  The iodine cell is used only for precision studies, for rest of the observations, by default cell is retracted from the beam path. The absorption lines of the iodine are imprinted on the stellar continuum during the cell-in position. The controller hardware and iodine cell assembly are integrated with the main spectrograph to facilitate regular observations.

Our goal is to expand the science capabilities of the existing spectrograph along with its regular observations. The  iodine technique that we describe in the next section can be used for characterization the hot Jupiter with velocity amplitudes up to few 100~$\textrm{ms}^{-1}$.

\section{Stellar observations with Iodine cell}
\label{sect:procedure}
\subsection{Observational strategy} 
The technique we employed is used to track and remove the instrumental shifts from the stellar observations. For RV studies, all the star observations are taken with iodine cell in the beam path. Two sets of observations are used in the analysis:
\begin{itemize}

\item \textbf{RV-star}: The target star for the RV measurements. These observations are taken with the cell in the beam path, thus producing composite spectra of iodine and stellar absorption lines.

\item \textbf{Ref-star}: A rapidly rotating hot star (spectral type O, A or B) without any spectral features in the wavelength range of Iodine (5000-6000~\AA). The iodine absorption lines are superimposed on the Ref-star continuum. 
\end{itemize}

 The RV-star observations are taken continuously and are bracketed with Ref-star observations. This approach is similar to the bracketing technique for UVES spectrograph that used attached (closest possible) Th-Ar observational frame for wavelength calibration \cite{super_calibration,uves}.

\begin{figure}
   \begin{center}
   \begin{tabular}{c} 
   \includegraphics[scale = 0.5]{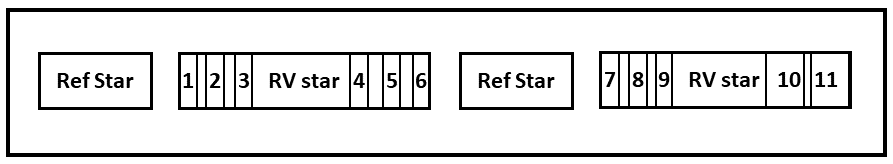}
	\end{tabular}
	\end{center}
   \caption
   { \label{fig:observationalflow} 
Proposed sequence of the RV and Ref-star observations. The Ref-stars observations are bracketed in between the sequence of RV-star observations. The closest Ref-star is used to correct for the instrumental shifts in the RV-star. The ideal case would be to take Ref-star observations after every RV-star, but the algorithm works under the assumption that the spectrograph is stable within a time of 30 minutes for reference. 
}
\end{figure} 

 In our approach, we use the iodine lines from the closest possible Ref-star as a reference for determining the instrumental shifts. The algorithm steps are explained with the help of an example illustrated in Figure \ref{fig:observationalflow}. In order to analyze the instrumental shifts, the first six RV-star observations are divided into two slots. The iodine lines in the first three RV-star (1,2,3) observations are compared with the position of the iodine lines in the first Ref-star. Similarly, the iodine lines in the next three RV-star observations (4,5,6) are compared with that of second Ref-star. This second Ref-star is shown in Figure \ref{fig:observationalflow} is the same Ref-star taken at a later point of time (usually at every 30 minutes interval). Correspondingly, the next set of observations (7,8,9) of RV-star are compared with closest possible Ref-star (second Ref-star) for instrumental shifts.  

Since the observations are taken with the cell in the beam path, the iodine lines that are common for both Ref-star and RV-star are used for analyzing the instrumental shifts.  The Ref-star has iodine lines and the RV-star has both star and iodine features. The instrument induced shifts are common for both stellar and iodine lines. The iodine lines in the Ref-star are used as a reference to measure the instrumental shifts registered in iodine lines of the RV-star. The spectra of RV-star is then corrected to take out the instrumental effect. Finally, all the instrumental shift that is induced in the iodine lines with respect to the first Ref-star observation is corrected. These shifts are induced due to the instrument/environmental variations and they are removed from the resulting RV calculations of the star.

\subsection{Instrumental shift removal algorithm}

We used IRAF for basic data reduction procedures such as bad pixel removal, scattered light subtraction, bias corrections, aperture extraction, flat fielding, and wavelength calibration. A Th-Ar lamp spectrum is used as a primary wavelength calibration source. The spectra are normalized in IRAF. The wavelength calibrated and normalized spectra are used as input to the instrument shift removal algorithm which is developed in Python. The main steps in the algorithm are as follows: 

\begin{enumerate}
\item \textbf{Interpolation of wavelength chunks}: 
The Iodine region spectral window  (5000-6000~\AA) is divided into smaller wavelength chunks of 5 \AA~ each\cite{marcy}. Each chunk is interpolated to obtain accurate sub-centroiding of the pixel \cite{interpolation}. 

\item \textbf{Instrumental shift analysis from closest (in time) Ref-star}: The RV-star (star + iodine) spectra is first compared with the closest Ref-star (iodine). The top panel of Figure \ref{fig:shift_corrected} shows the relative displacement in the iodine features between an RV and a Ref-star spectrum. We used the cross-correlation method to compute the spectral shift in each wavelength chunk \cite{correlation}.  A second order polynomial was fitted to the output of cross-correlation function to determine the spectral shift.

\item \textbf{Shifting of RV-star to the closest Ref-star spectra}: Based on the positional offset obtained from the cross-correlation, the RV-star spectrum is shifted to the Ref-star position as shown in the bottom panel of Figure \ref{fig:shift_corrected}. By taking out the instrumental shift this way the RV-star spectra is now well matched with the spectra of Ref-star nearest in time.

  \begin{figure}
   \begin{center}
   \begin{tabular}{c} 
   \includegraphics[scale = 0.4]{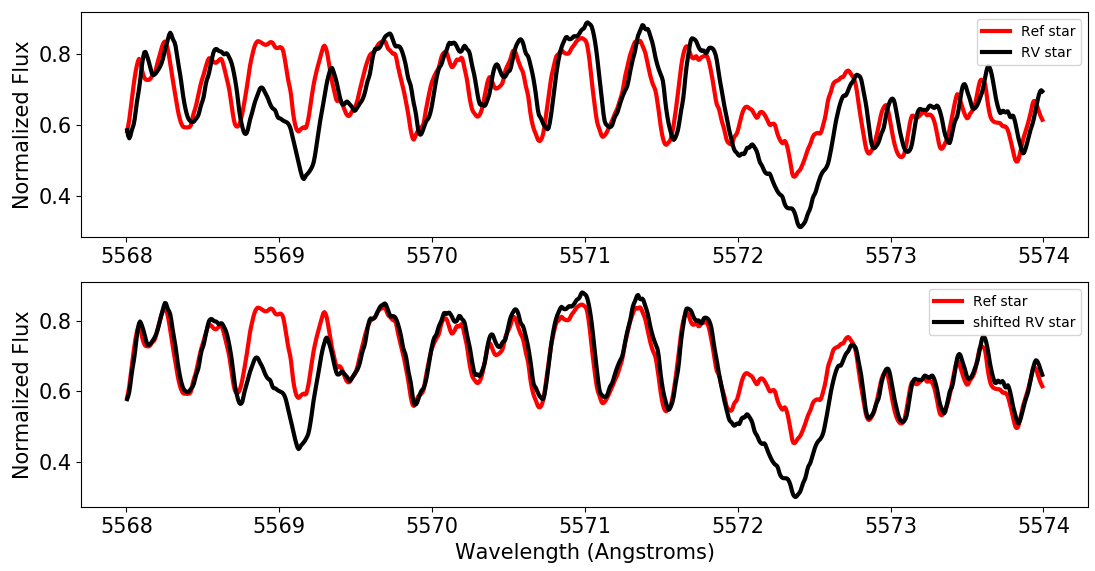}
	\end{tabular}
	\end{center}
   \caption
   { \label{fig:shift_corrected} 
Ref and RV-star observations made with iodine cell. The top panel of the image shows the spectra of Ref-star (marked in red) and RV-star (marked in black) taken with the iodine cell in the beam path. The Ref-star has no pronounced spectral features of its own and all the lines are seen are from the iodine gas. RV-star has both stellar lines and iodine features (e.g. 5569~\AA~ and 5572.5~\AA~). Iodine features are denser compared to stellar absorption lines. Horizontal offset between two spectra seen in the top panel is due to instrumental drift. The RV shift in target star is corrected in the bottom panel. }
   \end{figure} 
 
\item \textbf{Instrumental shift with respect to the first Ref-star }: The instrumental shift removed from RV-star spectra in step 3 is with respect to the closest (in time) Ref-star observations. All  Ref-star observations are correlated with the first spectral frame of  Ref-star. The RV-star spectra are shifted again with respect to the shifts calculated from the Ref-star. Later, all the RV-stars are brought to the first Ref-star position by eliminating the instrumental fluctuations. 

\item \textbf{Final RV of the star} : The RV-star observations are corrected for instrumental fluctuations. We use the first RV-star observation as template and cross-correlate all the instrument shift-corrected RV-star observations with respect to the template.  In order to find the shift in the wavelength calibrated spectra from the result of correlation, a second-order polynomial is fitted. The obtained wavelength shift is used to calculate the radial velocity shift of the RV-star with respect to the first (reference) RV-star. An average of all the wavelength chunks is obtained for the final relative RV shift. The standard deviation of all the wavelength chunks is taken as the error in the RV analysis.

\end{enumerate}

\section{Preliminary results}
\label{sect:results}

In order to validate our approach, we have observed a well-studied planet-hosting  star, tau Bootis \cite{tauboo1,tauboo}. The details of the star and the observations taken with the spectrograph at VBT are given in  Table \ref{tab:Ref_RV}. The observational log is given in Table \ref{tab:log}.

\begin{table}
\begin{center}

\begin{tabular}{|l|l|l|l|}
\hline
\textbf{RV-star Parameter} & \textbf{Value} & \textbf{Ref-star Parameter} & \textbf{Value} \\ \hline
Star Name                  & tau Bootis        & Star Name                   & HR5191         \\
Apparent magnitude (V)  & 4.5            & Apparent magnitude (V)   & 1.86           \\
Spectral type              & F7 V           & Spectral type               & B3V B          \\
Average exposure time      & 12 minutes     & Average exposure time       & 2 minutes     \\  \hline
\end{tabular}
\end{center}
\caption
   { \label{tab:Ref_RV}
Observational details of the target RV-star and Ref-star.  } 

\end{table}

\begin{figure}
   \begin{center}
   \begin{tabular}{c} 
   \includegraphics[scale = 0.6]{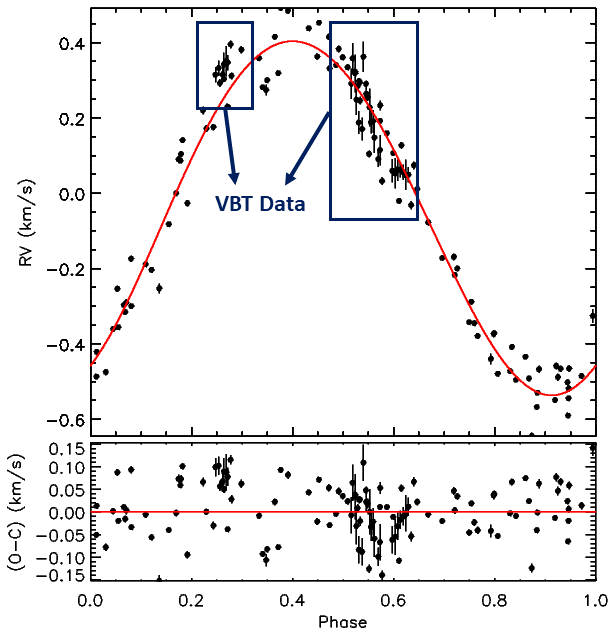}
	\end{tabular}
	\end{center}
   \caption
   { \label{fig:RV_one} 
   The Phase-folded plot of the tau Bootis obtained from the VBT spectrograph data and the archival data of Lick-Hamilton spectrograph \cite{RV_data}. The proposed algorithm in Section 3 was used to obtained RV of the star from the observations.}
   \end{figure}

We have used tau Bootis observations to test our algorithm and compared it with the data taken (available online at NASA's Exoplanet Archive \footnote[1]{https://exoplanetarchive.ipac.caltech.edu/}) with Lick-Hamilton spectrograph \cite{RV_data}.  The echelle spectrograph at VBT is equipped with the iodine absorption cell from March 2017. We have used the cell for observations during some of the nights from April 2017 - 2018. Various stars were observed during this course of the nights. The observational site experiences monsoon twice over the year. Though the sky conditions were not that favorable all the times, we have obtained some useful data to verify the efficacy of our algorithm. We have obtained a total of 32 observations with tau Bootis star observed with the iodine cell.  Figure \ref{fig:RV_one} shows the phase-folded analysis of star with our RV data along with the archival data using RVfit code in IDL \cite{rvfit}. The fitting parameters obtained from the phase folding analysis are given in Table \ref{tab:fit_details}. The fitting shows that the procedure we adopted gives results that are reasonably close to those obtained with precision RV spectrograph.

\begin{table}
\begin{center}

\begin{tabular}{|l|l|l|}
\hline
\textbf{Parameter}    & \textbf{Actual value} & \textbf{Obtained value}    \\ \hline
Orbital Period (days) & 3.31249               & 3.31241   \\
Tperi (JD)            & 2450529.20000         & 2450528.93764$\pm$ 0.00044 \\
Eccentricity          & 0.07870               & 0.03670$\pm$0.00056        \\ 
$K_1$ ($\textrm{kms}^{-1} $)         & 469.59 $\pm$ 14.86    & 470.302$\pm$0.00090        \\ \hline
\end{tabular}
\end{center}
\caption
   { \label{tab:fit_details} 
Fit parameters obtained from the RVfit code are given in the obtained values column. These values are compared with the published data for the star.}

\end{table}

\begin{table}
\begin{center}

\begin{tabular}{|l|l|l|}
\hline
\textbf{JD (2000)}  & \textbf{Radial velocity ($\textrm{kms}^{-1} $)} & \textbf{Uncertainity ($\textrm{kms}^{-1} $)} \\ \hline
2457871.1843 & 0.05995                         & 0.03                         \\
2457871.2022 & 0.05270                         & 0.04                         \\
2457871.2186 & 0.06292                         & 0.03                         \\
2457871.2366 & 0.06267                         & 0.02                         \\
2457871.2545 & 0.05602                         & 0.02                         \\
2457871.2715 & 0.04926                         & 0.04                         \\ 
2457871.2892 & 0.04884                         & 0.02                         \\
2458181.3911 & 0.31474                         & 0.02                         \\
2458181.4122 & 0.33236                         & 0.02                         \\
2458181.4205 & 0.29238                         & 0.01                         \\
2458181.4379 & 0.31520                         & 0.02                         \\
2458181.4463 & 0.34261                         & 0.03                         \\
2458181.4626 & 0.35096                         & 0.04                         \\
2458181.4719 & 0.34691                         & 0.02                         \\
2458181.49   & 0.39545                         & 0.01                         \\
2458182.2818 & 0.29121                         & 0.04                         \\
2458182.2902 & 0.35849                         & 0.04                         \\
2458182.307  & 0.31847                         & 0.05                         \\
2458182.3153 & 0.24898                         & 0.04                         \\
2458182.331  & 0.18770                         & 0.03                         \\
2458182.3392 & 0.24593                         & 0.01                         \\
2458182.3527 & 0.17042                         & 0.03                         \\
2458182.3615 & 0.36236                         & 0.04                         \\
2458182.3797 & 0.26432                         & 0.02                         \\
2458182.3881 & 0.25536                         & 0.01                         \\
2458182.4023 & 0.22777                         & 0.05                         \\
2458182.4106 & 0.18878                         & 0.03                         \\
2458182.425  & 0.19093                         & 0.03                         \\
2458182.4333 & 0.14799                         & 0.04                         \\
2458182.4584 & 0.09066                         & 0.02                         \\
2458182.4697 & 0.11465                         & 0.04                         \\
2458182.4834 & 0.03204                         & 0.01                         \\
 \hline
\end{tabular}
\end{center}
\caption
   { \label{tab:log} 
Julian date of observations and corresponding radial velocities obtained for tau Bootis. }

\end{table}
\section{Discussion}
\label{sect:disc}

\subsection{Impact of correlating Ref and RV-star}

We analyzed the impact of correlating Ref-star (iodine lines) with RV-star (product of iodine and star lines) on the final radial velocity obtained in the following ways: 

\begin{itemize}
\item \textit{Synthetic spectrum of the star and high-resolution template spectrum of iodine:}

The synthetic spectra of the star and the high-resolution template spectrum of the iodine cell is used to evaluate the errors obtained due to cross-correlation. The iodine lines in the template spectrum are given a known amount of shift to verify if the algorithm predicts and corrects for the shift in the iodine lines. Ideally, the iodine lines are to be brought back to the reference position with zero shift. However, as the iodine spectrum gets multiplied by the stellar lines, the correlation of Ref-star spectrum with the RV-star to detect the shift in the iodine lines might not be accurate to a scale of sub $\textrm{ms}^{-1} $. The shift given to the template iodine spectrum in the simulation is evaluated for correction.  The error obtained in the analysis of mean radial velocity is 12.82 $\textrm{ms}^{-1} $.

 Similarly, a star with small rotational velocity is also analyzed to compare the effect on the correlation result. In this scenario, we have obtained the template spectrum of 51 Pegasi (V sin i = 2 $\textrm{kms}^{-1} $). The error obtained in the mean radial velocity is 9.8 $\textrm{ms}^{-1} $. 
 
\item  \textit{Analysing the effect of star lines in terms of depth of the lines and blending of star lines:}

One of the RV-star is analyzed for radial velocity value from individual chunks in comparison with the final mean velocity. The RV values in the chunks above the standard deviation from the mean position are compared with stellar absorption lines in the obtained spectra. These deviations would arise from the blending of star lines and due to the depth of the star lines as shown in Figure \ref{fig:RV_chunks}. We have compared the tau Bootis spectrum taken with the spectrograph with the RV obtained using the algorithm discussed above. 

As shown in Figure \ref{fig:RV_chunks}, the last chunk has a broadened absorption feature of the star with a depth of around 0.65  from the normalized value. This has led to the deviation of the corresponding RV obtained from the mean position. The obtained RV values are within the deviation from the mean position where there is a lack of strong stellar features. To further improve the accuracy of the technique, the star lines can be masked and the shift in the iodine lines between RV and Ref-star is to be estimated.

  \begin{figure}
   \begin{center}
   \begin{tabular}{c} 
   \includegraphics[scale = 0.40]{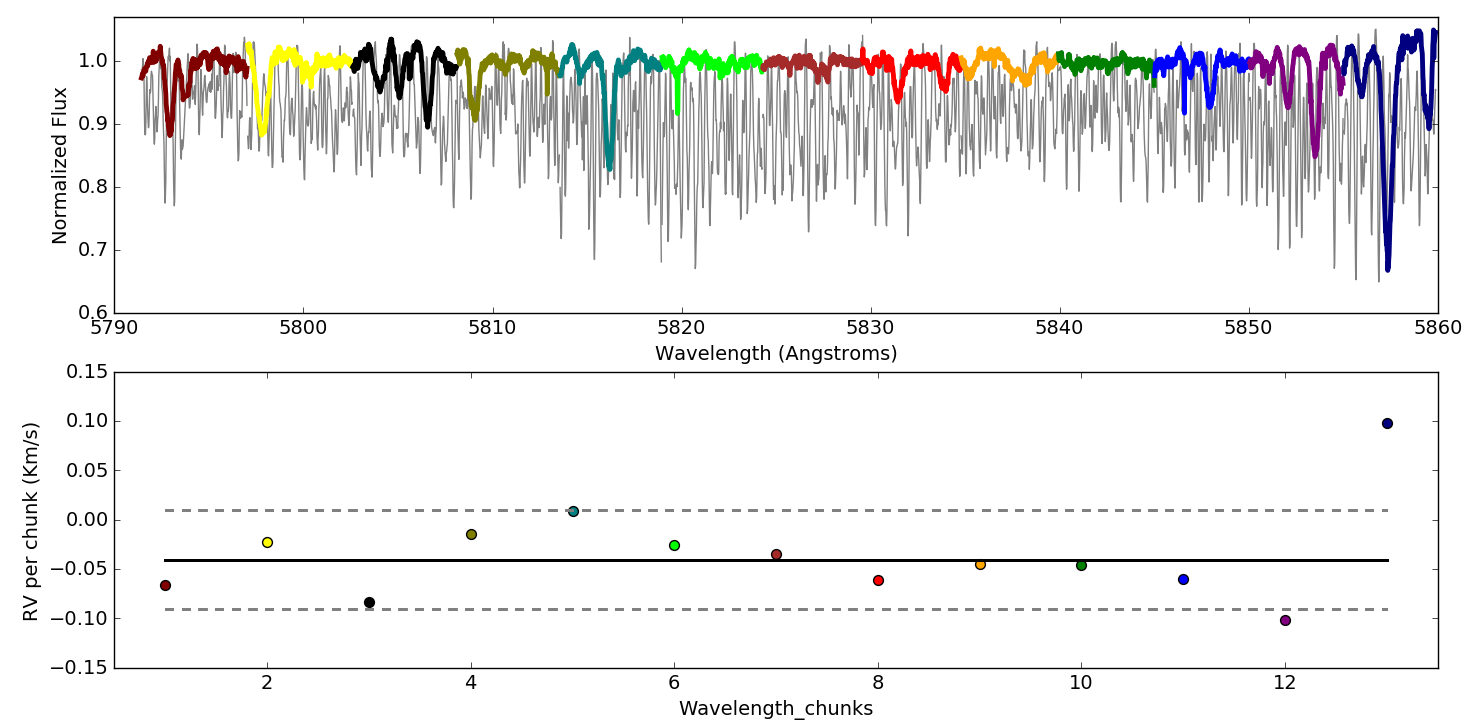}
	\end{tabular}
	\end{center}
   \caption
   { \label{fig:RV_chunks} 
Tau Bootis spectral features are compared with the RV shift determined  (along with deviation from the mean shift) from individual wavelength chunks. The depths of the stellar lines and blending is analyzed for the impact on the RV error. The top panel shows a Tau Bootis spectrum taken with the spectrograph.  Each color represents an individual wavelength chunk along within a particular order. The spectrum represented in the gray is that of Ref-star in the same order. Bottom panel shows the RV values obtained from cross-correlation for individual wavelength chunk. The solid line is the mean, while the dashed lines represent $1\sigma$ departure from the mean.}
   \end{figure} 

\end{itemize}

\subsection{Comparison with bracketing ThAr approach}

In the case of bracketing ThAr approach, frequent ThAr lamps are used to correct the wavelength calibration scale and estimate the shifts in the calibration lamp lines before and after the stellar observations. However, this cannot be used to track the instrumental shifts during observations directly like the case of Iodine absorption cell technique discussed. This technique replaces the frequent ThAr exposures with Ref-star exposures before and after RV-star observations. These Ref stars are chosen as very bright stars and are available within or in the nearby field of the RV star observations to avoid observational time loss. 

The shortcoming of the algorithm, however, lies in the introduction of iodine cell in the beam path, thus reducing the transmission efficiency compared to bracketing ThAr approach. Apart from that, this technique requires the Ref-star observations at an interval of 15 - 30 minutes with the spectrograph. In our case, the limiting magnitude of the target RV-star was V < 7, which also include non-optimal reflectivity of the telescope and degradation of anti-reflection coating on the cell windows. 

The reliability of the algorithm lies in the frequency of Ref-star observations.  In case of the Echelle spectrograph at VBT, the spectrograph is stable to a level of $\sim$40~$\textrm{ms}^{-1} $ within the time-frame of 15 - 30 minutes. Thus, there is an assumption that the instrument remains stable in that short duration and also that PSF is invariant within smaller wavelength chunk (typically 2--5~\AA). This time interval of bracketing observations is to be decided based on the stability of the spectrograph.

\section{Conclusion}
\label{sect:concl}

The traditional iodine technique uses complex modelling to measure the relative shift in the Doppler signal in star's spectra \cite{PSF}.  Additionally, a high-resolution Fourier Transform template of the iodine spectra are also required to be maintained as part of the RV extraction procedure. We have proposed an alternative but a generic approach of using iodine absorption cell to eliminate the instrumental fluctuations from the RV observations of the star. Our algorithm is not substituting for the original forward modelling approach developed by Marcy and Butler \cite{marcy} but it can be used for general purpose spectrographs with inherent limitations like movable components and environmental fluctuations.

The main limitation of our spectrograph is the movable grating leading to a lack of repeatable zero-point. We have developed a Zemax model to predict and correct for the zero-position of the instruments \cite{Zemax}. This is to be upgraded and then the validation is to be done to compare the improvement in the results. Thus, we can use the Echelle spectrograph operating at VBT for precision follow up studies along with the other high-resolution spectroscopic observations. Our target is not to build a dedicated instrument for precision radial velocity measurements, but to upgrade the spectrograph to achieve a moderate level precision between 10-100~$\textrm{ms}^{-1}$ range, useful for RV follow up studies that are not very challenging. 
  
\section{Acknowledgements}

Authors would like to thank the technical and observing staff at Vainu Bappu Telescope. We also acknowledge the valuable support received from Dr. Sivarani Thirupathi, Mr. Sagayanathan. K, Mr. Anbazhagan. P, Mr. Ravi. K, Mr. Naveen Kumar. R, Mr. Pandiyarajan. B during different phases of this work. This research made use of NASA's exo-planet archival data, which is operated by California Institute of Technology. Finally, we would also like to thank referees for their critical review and useful feedback that helped us to improve the manuscript.


\bibliography{report}   

\begin{thebibliography}{10}

\bibitem{first_exo}
M.~Mayor and D.~Queloz, ``A jupiter-mass companion to a solar-type star,'' {\em
  Nature} {\bf 378}(6555), 355  (1995).

\bibitem{marcy}
G.~W. Marcy and R.~P. Butler, ``Precision radial velocities with an iodine
  absorption cell,'' {\em Publications of the Astronomical Society of the
  Pacific} {\bf 104}(674), 270  (1992).

\bibitem{Pepe}
F.~Pepe, C.~Lovis, D.~Segransan, {\em et~al.}, ``The harps search for
  earth-like planets in the habitable zone-i. very low-mass planets around hd
  20794, hd 85512, and hd 192310,'' {\em Astronomy \& Astrophysics} {\bf 534},
  A58  (2011).

\bibitem{onems}
D.~Queloz, M.~Mayor, S.~Udry, {\em et~al.}, ``From coralie to harps. the way
  towards 1 m s-1 precision doppler measurements,'' {\em The Messenger} {\bf
  105}, 1--7  (2001).

\bibitem{VBTechelle}
N.~K. Rao, S.~Sriram, K.~Jayakumar, {\em et~al.}, ``High resolution stellar
  spectroscopy with vbt echelle spectrometer,'' {\em Journal of Astrophysics
  and Astronomy} {\bf 26}(2-3), 331--338  (2005).

\bibitem{uves}
P.~Molaro, S.~Levshakov, S.~Monai, {\em et~al.}, ``Uves radial velocity
  accuracy from asteroid observations-i. implications for fine structure
  constant variability,'' {\em Astronomy \& Astrophysics} {\bf 481}(2),
  559--569  (2008).

\bibitem{keck}
K.~Griest, J.~B. Whitmore, A.~M. Wolfe, {\em et~al.}, ``Wavelength accuracy of
  the keck hires spectrograph and measuring changes in the fine structure
  constant,'' {\em The Astrophysical Journal} {\bf 708}(1), 158  (2009).

\bibitem{stability_VBT}
S.~Chamarthi, R.~K. Banyal, S.~Sriram, {\em et~al.}, ``Stability analysis of
  vbt echelle spectrograph for precise radial velocity measurements,'' {\em
  Journal of Optics} , 1--7  (2017).

\bibitem{whitepaper}
P.~Plavchan, D.~Latham, S.~Gaudi, {\em et~al.}, ``Radial velocity prospects
  current and future: A white paper report prepared by the study analysis group
  8 for the exoplanet program analysis group (exopag),'' {\em arXiv preprint
  arXiv:1503.01770}   (2015).

\bibitem{autoguider}
I.~Boisse, F.~Bouchy, B.~Chazelas, {\em et~al.}, ``Consequences of spectrograph
  illumination for the accuracy of radial-velocimetry,'' in {\em EPJ Web of
  conferences},   {\bf 16}, 02003, EDP Sciences  (2011).

\bibitem{tauboo1}
Y.~Takeda, B.~Sato, E.~Kambe, {\em et~al.}, ``Iodine-cell spectroscopy at
  okayama astrophysical observatory: First results,'' {\em Publications of the
  Astronomical Society of Japan} {\bf 54}(1), 113--120  (2002).

\bibitem{tauboo}
T.~G. Kaye, S.~Vanaverbeke, and J.~Innis, ``High-precision radial-velocity
  measurement with a small telescope: Detection of the tau bootis exoplanet,''
  {\em arXiv preprint astro-ph/0609468}   (2006).

\bibitem{rvfit}
R.~Iglesias-Marzoa, M.~L{\'o}pez-Morales, and M.~J.~A. Morales, ``The rvfit
  code: a detailed adaptive simulated annealing code for fitting binaries and
  exoplanets radial velocities,'' {\em Publications of the Astronomical Society
  of the Pacific} {\bf 127}(952), 567  (2015).

\bibitem{systemic}
S.~Meschiari, A.~S. Wolf, E.~Rivera, {\em et~al.}, ``Systemic: a testbed for
  characterizing the detection of extrasolar planets. i. the systemic console
  package,'' {\em Publications of the Astronomical Society of the Pacific} {\bf
  121}(883), 1016  (2009).

\bibitem{PSF_vbt}
S.~Chamarthi, R.~K. Banyal, and S.~Sriram, ``Estimation of asymmetries in point
  spread function for the echelle spectrograph operating at vainu bappu
  telescope for high precision radial velocity studies,'' in {\em Ground-based
  and Airborne Instrumentation for Astronomy VII},   {\bf 10702}, 1070275,
  International Society for Optics and Photonics  (2018).

\bibitem{PSF}
J.~A. Valenti, R.~P. Butler, and G.~W. Marcy, ``Determining spectrometer
  instrumental profiles using fts reference spectra,'' {\em Publications of the
  Astronomical Society of the Pacific} {\bf 107}(716), 966  (1995).

\bibitem{Zemax}
S.~Chamarthi, R.~K. Banyal, and S.~Sriram, ``Sensitivity studies of fibre fed
  echelle spectrograph at vbt for precision radial velocity measurements,''
  {\em In Preparation}   (2018).

\end{thebibliography}
\bibliographystyle{spiejour}   

\listoffigures
\listoftables

\end{spacing}
\end{document}